\documentclass[aip,
rsi,
 amsmath,amssymb,
 reprint,%
]{revtex4-1}

\usepackage{etoolbox}
\usepackage[american]{babel}
\let\oldselectlanguage\selectlanguage
\renewcommand{\selectlanguage}[1]{%
  \ifstrequal{#1}{en}{\oldselectlanguage{english}}{%
    \oldselectlanguage{#1}%
  }%
}
\usepackage{graphicx}
\usepackage{dcolumn}
\usepackage{bm}
\usepackage{siunitx}
\DeclareSIUnit\angstrom{\text{Å}}
\usepackage[utf8]{inputenc}
\usepackage[T1]{fontenc}
\usepackage{mathptmx}

\usepackage{xspace}
\newcommand{\xray}{X\nobreakdash-ray\xspace}
\newcommand{\xrays}{X\nobreakdash-rays\xspace}

\makeatletter
\def\@email#1#2{%
 \endgroup
 \patchcmd{\titleblock@produce}
  {\frontmatter@RRAPformat}
  {\frontmatter@RRAPformat{\produce@RRAP{*#1\href{mailto:#2}{#2}}}\frontmatter@RRAPformat}
  {}{}
}%
\makeatother
\begin{document}

\title[Automated laboratory \xray diffractometer and fluorescence spectrometer for high-throughput materials characterization]{Automated laboratory \xray diffractometer and fluorescence spectrometer for high-throughput materials characterization}
\author{Hyun Sang Park }
 \email{hpark108@jh.edu}
\affiliation{ 
Johns Hopkins University
}
\author{Timothy Long}
 \email{tlong25@jh.edu}
\affiliation{ 
Johns Hopkins University
}%
\author{Michael Wall}%
 \email{mwall11@jh.edu}
\affiliation{ 
Johns Hopkins University
}%
\author{Alexander deJong}%
 \email{adejong2@jh.edu}
\affiliation{ 
Johns Hopkins University
}%
\author{Ali Rachidi}%
 \email{arachid1@jh.edu}
\affiliation{ 
Johns Hopkins University
}%
\author{Kacper Kowalik}%
 \email{kowalikk@illinois.edu}
\affiliation{ 
NCSA, University of Illinois at Urbana-Champaign
}%

\author{Rohit Berlia}
\affiliation{ 
Johns Hopkins University
}%

\author{David Elbert}%
 \email{elbert@jhu.edu}
\affiliation{ 
Johns Hopkins University
}%

\author{Timothy P. Weihs}
\affiliation{ 
Johns Hopkins University
}%

\author{Robert Drake}%
 \email{rdrake@protoxrd.com}
\affiliation{ 
PROTO Manufacturing Ltd
}%
\author{Todd C. Hufnagel}%
 \email{hufnagel@jhu.edu}
\affiliation{ 
Johns Hopkins University
}%

\date{\today}

\begin{abstract}
The increasing importance of artificial intelligence and machine learning in materials research has created demand for automated, high-throughput characterization techniques capable of rapidly generating large data sets. We describe here a new instrument for simultaneous \xray diffraction and \xray fluorescence spectroscopy, optimized for high-throughput studies of combinatorial specimens. A bright, focused, high-energy \xray beam (\qty{24}{keV}) combined with a pixel array area detector allows spatially-resolved ($\sim$\qty{200}{\micro\meter}) transmission diffraction measurements through thick ($>$\qty{100}{\micro\meter}) specimens of structural metals with exposure times as short as \qty{1}{s}. Simultaneously, a silicon drift detector records \xray fluorescence from the specimen for spatially-resolved measurement of composition. Specimen handling is fully automated, with a robot inside the \xray enclosure manipulating the sample for measurements at different locations. Data orchestration is also automated, with data streamed off the instrument and processed autonomously. In this paper we assess the performance of the instrument in terms of throughput, resolution, and signal-to-noise ratio, and provide an example of its capabilities through a combinatorial study of Cu-Ti alloys to demonstrate rapid data set creation.

\end{abstract}
\maketitle
\section{\label{introduction}Introduction}
The recent rise in data-intensive machine-learning (ML) models in materials research has put a spotlight on automated high-throughput laboratories, which improve the efficiency and reduce the cost of generating large data sets~\cite{adapa_rapid_2025, talley_research_2021, burger_mobile_2020}. Of particular importance are combinatorial high-throughput (CHT) techniques in which there is a gradient in chemistry, microstructure, or both across a single specimen, with spatially-resolved measurements used to obtain data as a function of position~\cite{gregoire_high-throughput_2014, maruyama_high-throughput_2020, zhao_combinatorial_2006, maier_combinatorial_2007,potyrailo_combinatorial_2011, ludwig_discovery_2019}. Most CHT studies to date have focused on thin-film combinatorial libraries, largely because it is convenient to produce appropriate specimens by vapor deposition of thin films on substrates. Such samples may be well-suited for studies of functional properties, but are not ideal for studies of mechanical properties of bulk materials for two reasons. First, vapor-deposited thin films tend to have strong crystallographic texture and fine grain sizes that are not representative of bulk materials and are not easily varied by thermomechanical processing. Second, some mechanical properties and behaviors of interest (including yield strength and ductility) are strongly influenced by the size of the specimen itself, and studies conducted on thin films often do not carry over to bulk materials.

As a result, there is a need for high-throughput characterization techniques for bulk structural metals, including the capability to assess microstructure~\cite{miracle_emerging_2021}. Electron backscatter diffraction (EBSD) would seem to be an obvious choice because it provides a wealth of information about microstructure, including grain size distributions, crystallographic texture, and local grain misorientations~\cite{dingley_phase_2009, shekhar_electron_2022, schwartz_electron_2009}. But EBSD requires time-consuming surface preparation, the details of which are different for different materials and therefore may be problematic across a combinatorial sample (although we do note that progress is being made using techniques such as dry electropolishing\cite{fowler_high-throughput_2024, farrer_ebsd_2000}). A related issue is that the structural information in EBSD comes from a shallow region near the surface (\qtyrange{10}{50}{\nano\meter}, depending on material), making it insensitive to any through-thickness variations in microstructure. These traits prevent EBSD from probing bulk structures without time-consuming serial sectioning. Finally, the large specimen tilt required for EBSD is problematic for large combinatorial samples, which may interfere with the microscope pole piece.

Transmission high-energy \xray diffraction (XRD) offers an attractive alternative for high-throughput characterization of bulk metals and alloys. Little or no surface preparation is required, the measurement inherently averages over the projected thickness of the specimen, and large specimens can be easily examined. Although the microstructural information is not as rich as that provided by EBSD, it readily provides quantitative phase analysis and lattice parameters, and a 2D diffraction pattern encodes information about crystallographic texture and grain size as well.~\cite{he_miscellaneous_2018,he_texture_2018} 

Although high-throughput studies using XRD in a reflection geometry using both synchrotron~\cite{gregoire_high-throughput_2014, maruyama_high-throughput_2020} and laboratory sources~\cite{yotsumoto_autonomous_2024} have been reported, this geometry is somewhat limiting for combinatorial specimens due to spreading of the incident beam across the surface, which reduces the spatial resolution. For this reason a normal-incidence transmission geometry is preferred, which has the additional benefit of sampling the bulk (through-thickness) microstructure. This requires an \xray beam with high energy (for penetrating thick specimens) and high brightness (for spatial resolution), and several studies using synchrotron radiation have been reported ~\cite{kohara_high-energy_2003, vavrik_transmission_2025, reinle-schmitt_exploring_2023, he_synchrotron_2021}. But recent advances in source and detector technology are now making techniques previously limited to synchrotron sources possible in ordinary laboratories.~\cite{oh_taking_2025} 

In this paper we describe an instrument for automated high-throughput transmission \xray diffraction studies of materials, especially bulk structural metals. Because it is capable of simultaneous measurement of elemental composition using \xray fluorescence spectroscopy (XRF), we refer to it as MAXIMA, for ``Multi-modal Automated \xray Investigation of Materials.'' Its key components are a high-energy, high-brightness \xray source, focusing incident beam optics, a CdTe pixel array detector for XRD, a silicon drift detector (SDD) for XRF, and robotic sample handling. Another key feature is a sophisticated data framework, including autonomous data streaming and analysis. 

\section{\label{Instruments}Instrument configuration}
\subsection{\label{AIMD} AIMD-L}
Before discussing MAXIMA itself, we briefly describe the laboratory of which it is part. The Artificial Intelligence for Materials Design Laboratory (AIMD-L) was designed for high-throughput characterization of the structure and properties of structural metals and ceramics, especially in extreme environments. It comprises three major substations: microstructural characterization (MAXIMA), mechanical testing at relatively low loading rates (nanoindentation), and mechanical response due to shock loading using laser-driven microflyer impact. Physically, the systems are linked by a central conveyance for moving samples through the lab, with a Universal Robotics UR10 robot at each station to transfer samples to/from the experimental stations.

\subsection{\xray source and optics}
The primary requirement of the MAXIMA source is the ability to provide an intense, focused \xray\ beam with sufficiently high energy to penetrate bulk metallic specimens. We note the use of focusing optics requires a source with both high intensity and small size (\emph{i.e.} high brightness). In MAXIMA this is accomplished with an Excillum MetalJet E1+ with a liquid In-Sn-Ga alloy as the anode~\cite{noauthor_metaljet_nodate} which is capable of producing a source size as small as \qty{5}{\micro m} (though in practice we typically use a larger size to minimize cathode poisoning and increase source life expectancy).

For transmission measurements through structural metals, a high \xray energy is obviously desirable. Absorption in normal-incident transmission is described by the Beer-Lambert equation,
\begin{equation}
    \frac{I}{I_0} = \exp\left(-\mu t\right) = \exp\left(-\frac{t}{L_{\rm abs}}\right),
    \label{eq:intensity_absorption}
\end{equation}
where $I$ is the transmitted intensity, $I_0$ is the intensity of the initial beam, $t$ is the sample thickness, and $\mu$ is the linear absorption coefficient.\cite{cullity_elements_2014} The optimal thickness of a specimen for transmission XRD is one absorption length ($L_{\rm abs}=1/\mu$), which balances absorption of the \xray beam with the scattering volume: A sample of zero thickness would have zero absorption but also no material to scatter; an infinitely thick specimen would absorb all of the \xrays.\cite{nassar_structural_2016} 

\begin{figure}[tb]
    \centering
    \includegraphics[width=0.5\textwidth]{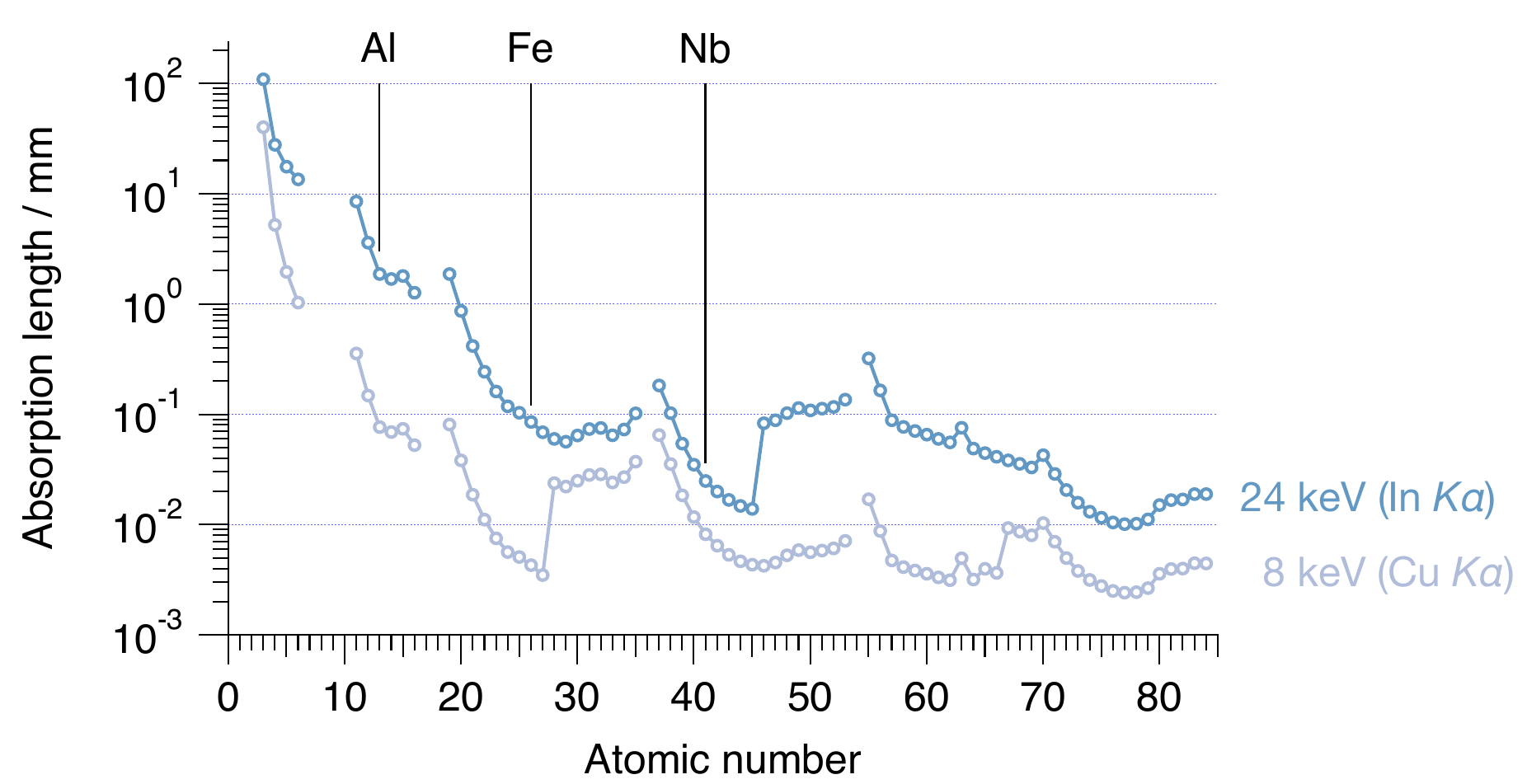}
    \caption{Absorption lengths of various elements for MAXIMA (In $K\alpha_1$, \qty{24.21}{keV}) and a standard laboratory source (Cu $K\alpha_1$, \qty{8.05}{keV}). The atomic numbers of three representative metals are indicated.
    }
    \label{fig:absorption lengths}
\end{figure}

For MAXIMA we use indium $K\alpha_1$ characteristic radiation, with an energy of \qty{24.21}{keV} (wavelength $\lambda$ = \qty{0.512}{\angstrom}). Figure \ref{fig:absorption lengths} shows $L_{\rm abs}$ for this energy for a wide range of elements, and we see that for first-row transition metals such as Fe the optimal sample thickness is $\sim$\qty{0.1}{mm}. It is still possible to make measurements on thicker (or thinner specimens) by counting longer; as a practical matter we find that count times are reasonable ($<$\qty{100}{s}, say) even for samples with $t\approx 5L_{\rm abs}$, as discussed below.

\begin{figure*}[htbp]
    \centering
    \includegraphics[width=\textwidth]{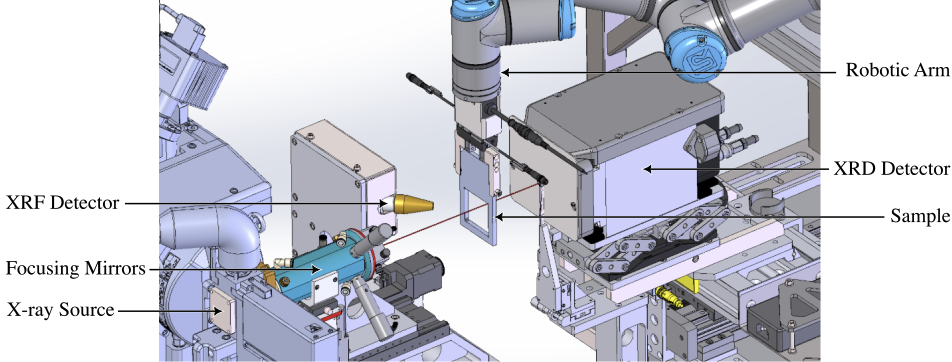}
    \caption{Schematic of MAXIMA performing XRD in transmission and XRF in reflection geometries. Red line depicts X-ray path relevant to transmission experiments.}
    \label{fig:Schematic}
\end{figure*}

Figure~\ref{fig:Schematic} shows the layout of the basic components of MAXIMA. \xrays emitted by the source are monochromatized and focused by orthogonal ellipsoidal graded multilayer mirrors (ExMontel-I-110/390/3.15). The instrument is designed for the \xrays to focus at the detector; with a convergence angle of approximately \qty{3.15}{\milli\radian}, the spot size at the nominal sample position is $\approx$ \qty{250}{\micro\meter}, which determines the spatial resolution (for combinatorial specimens, for example). With the source operating at its maximum rated power (\qty{1}{kW}) the nominal flux after the mirrors is $\sim$\qty{1e8}{ph.s^{-1}}.

\subsection{Detectors}
For \xray diffraction, MAXIMA uses a Eiger2 R CdTe 1M Dectris pixel array detector \cite{noauthor_dectris_nodate} with a pixel size of \qty{75}{\micro m}. The CdTe detection layer provides high efficiency for \qty{24}{keV} radiation. Because it is a photon-counting detector there is no readout noise or dark current, making it ideal for short exposures. Furthermore, it provides some energy discrimination to reject fluorescence from the specimen which would contribute background intensity. The Eiger2 detector is mounted on linear translation stages that permit up to \qty{500}{mm} travel parallel to, and \qty{76}{mm} perpendicular to the incident \xray beam. 

XRF measurements are done with a Fast Ultra high performance silicon drift detector \cite{noauthor_fast_nodate}. As shown in Fig.~\ref{fig:Schematic}, the SDD is placed on the upstream side of the specimen (to maximize intensity), nominally \qty{4}{cm} from the specimen surface and at an angle of $\sim$\qty{30}{\degree} with respect to the incident beam. It is also on a movable stage to adjust the geometry for different experiments.
It is important to note that MAXIMA was designed primarily for XRD, with XRF as a supplementary capability. In particular, the \xray source is monochromatic, significantly reducing the flux relative to a typical XRF source. In addition, the measurements are made in air, meaning that light elements ($Z<12$) cannot be detected. Still, the system is capable of quantifying major elements in most alloys, as demonstrated below.
\subsection{Sample handling}
In AIMD-L samples are delivered to MAXIMA via a belt-driven conveyance. When a measurement is to be made, an automatic safety door on the rear of the instrument opens, and a Universal Robotics UR10 robot transfers the sample from the conveyance to a holder inside the instrument. From there, an internal UR3e robot picks up the sample and holds it in position for measurements. The robot has a  nominal positional repeatability of $\pm$\qty{30}{\micro\meter}, enabling highly consistent and reproducible alignment of samples. The arm can also translate the sample in place, permitting us to make measurements in linear and grid arrays. The combination of internal and external robotics allows MAXIMA to run samples continuously without manual intervention. 

\subsection{Data streaming and workflows}
Achieving high throughput and integration with AI/ML algorithms requires sophisticated data handling. The Dectris detector provides high-speed readout to on-board storage; data are then streamed autonomously and consumed to Ceph object storage in our data center using OpenMSIStream \cite{eminizer_openmsistream_2023}, on an Apache Kafka backbone \cite{noauthor_apache_nodate}. Data are curated into our data portal using stream processing for metadata registration and managed data provenance.  Streaming ingest ensures that raw experimental results are available in real-time for downstream workflows. 

Data processing utilizes automated workflow orchestration using Dagster\cite{noauthor_dagster_nodate}. When new XRD data are streamed, the workflow performs a correction for detector tilt and calibrates the sample-to-detector distance. The workflow uses PyFAI\cite{kieffer_pyfai_2013, kieffer_pyfai_2013-1, kieffer_new_2020, ashiotis_fast_2015} to locate the beam center and azimuthally integrate the data to produce a 1D plot of intensity \emph{v} scattering vector magnitude ($q=4\pi\sin\theta/\lambda$ where $\theta$ is one-half of the scattering angle and $\lambda$ is the \xray wavelength). XRF data are similarly streamed  and analyzed using a PyMCA\cite{sole_multiplatform_2007} fundamental parameters pipeline. 

Processed data are written back to the data platform with concomitant metadata registration. All steps run inside containerized environments, enabling reproducible execution and isolation. Tasks are executed in parallel where possible to reduce overall latency and support high-throughput operation. The autonomous data handling and consistent handling of metadata mean that the data (and metadata) are immediately available for inspection and use by a human operator or an AI agent.

\section {Performance}
\label{sec:performance}
To demonstrate the performance of MAXIMA, we show results from two types of specimen: a \qty{1.5}{mm} thick NIST alumina line profile standard (SRM~1976c) and a set of vapor deposited combinatorial copper-titanium alloy foils. The Cu-Ti foils ($\sim$\qty{200}{\micro m} thick) were produced by magnetron sputtering using two targets of oxygen-free high conductivity (OFHC) copper (99.99\% purity) and one titanium target (99.95\% purity). Deposition was performed at room temperature and in an  Ar atmosphere of \qty{1.8}{\milli Torr} in a chamber with a base pressure of \qty{5e{-6}}{\milli Torr}. The sputtering power was \qty{800}{W} for the Cu and \qty{770}{W} for Ti. The substrates for the deposited foils were \qty{5}{in}$\times$\qty{12}{in} sheets of polished brass, masked with strips of \qty{3}{mm} wide kapton tape to produce individual \qty{40}{mm}$\times$\qty{40}{mm} samples. The substrates were affixed to a carousel which rotated at \qty{2.5}{Hz} during deposition, passing the substrates by each target in turn and thereby maintaining uniform thickness. A triangular shield on the Ti target produced a controlled composition gradient along the long dimension of the substrate ranging from nearly pure Cu (sample 1 below) to $\sim\qty{6.5}{at.\%}$ Ti (sample 6). After deposition the specimens were peeled from the substrates to produce free-standing foils for thermal processing and characterization. All samples were solutionized together by annealing in vacuum (\qty{2e{-6}}{Torr}) for \qty{10}{hr} at \qty{900}{\celsius} followed by oil quenching. All samples were analyzed in this as-produced state with no surface preparation. 

\subsection{Throughput}
\begin{figure}[tbp]
\centering
\includegraphics[width=0.45\textwidth]{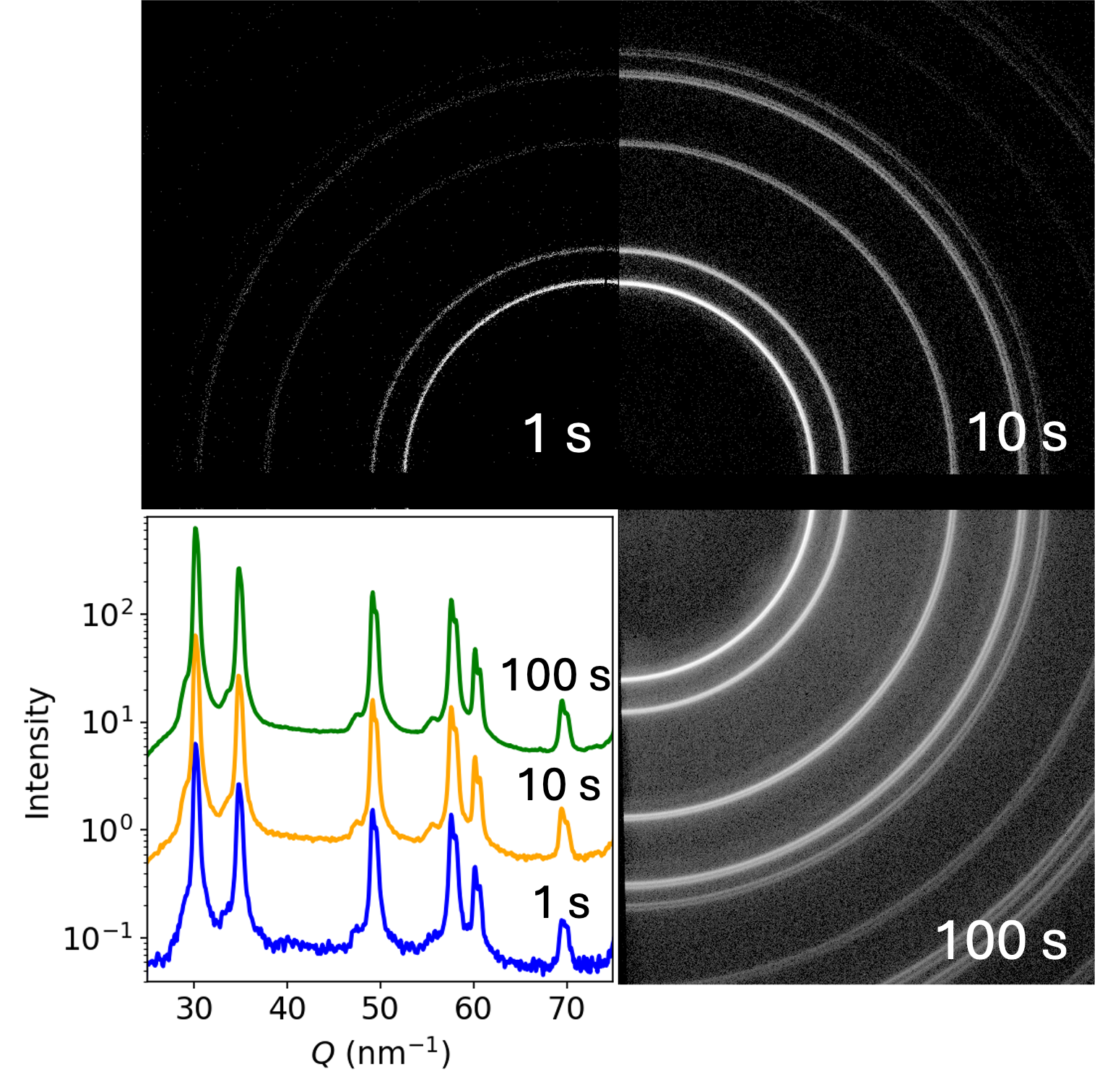}
    \caption{Diffraction patterns from a \qty{200}{\micro m} thick specimen of nearly pure Cu at $SDD$=\qty{63.7}{\milli\meter} from the detector for collection times from \qtyrange{1}{100}{s}.}
    \label{SNR with times}
\end{figure}

The single most important consideration in the design of the instrument was the desire to maximize throughput. Fig.~\ref{SNR with times} shows data for different collection times for a \qty{200}{\micro m} thick nearly pure Cu sample (taken from one of the combinatorial specimens). An absorption length for In K$\alpha$ radiation in Cu is about \qty{55}{\micro m}, so the sample thickness $t\simeq 3.6L_\mathrm{abs}$, which translates to a transmitted intensity of only about 2\% of the incident beam intensity. Despite the strong absorption, reasonable diffraction patterns were obtained even with collection times as short as \qty{1}{s}. Longer times serve to improve the counting statistics, which helps resolve small peaks and improves angular resolution based on sub-pixel splitting algorithm approaches\cite{he_miscellaneous_2018}. Increased scan times can also compensate for signal attenuation in even thicker samples. This enables consistent data quality when mechanical thinning is not feasible, such as with highly sensitive materials, hard materials, or mechanical characterization specimens with specific thickness requirements.

The overall throughput of MAXIMA is also enhanced by a high level of automation, the lack of surface preparation, and that the measurements are performed in air (as opposed to EBSD). We have recently demonstrated this by an automated series of experiments in which multiple samples were sequentially transferred from the AIMD-L conveyance system into MAXIMA, measured, and then returned to the conveyance. In this demonstration, MAXIMA collected 162 diffraction patterns from 6 different specimens in approximately 90 minutes. 

\subsection{XRD resolution}
\label{sec:resolution}
\begin{figure}[tbp]  \includegraphics[width=0.45\textwidth]{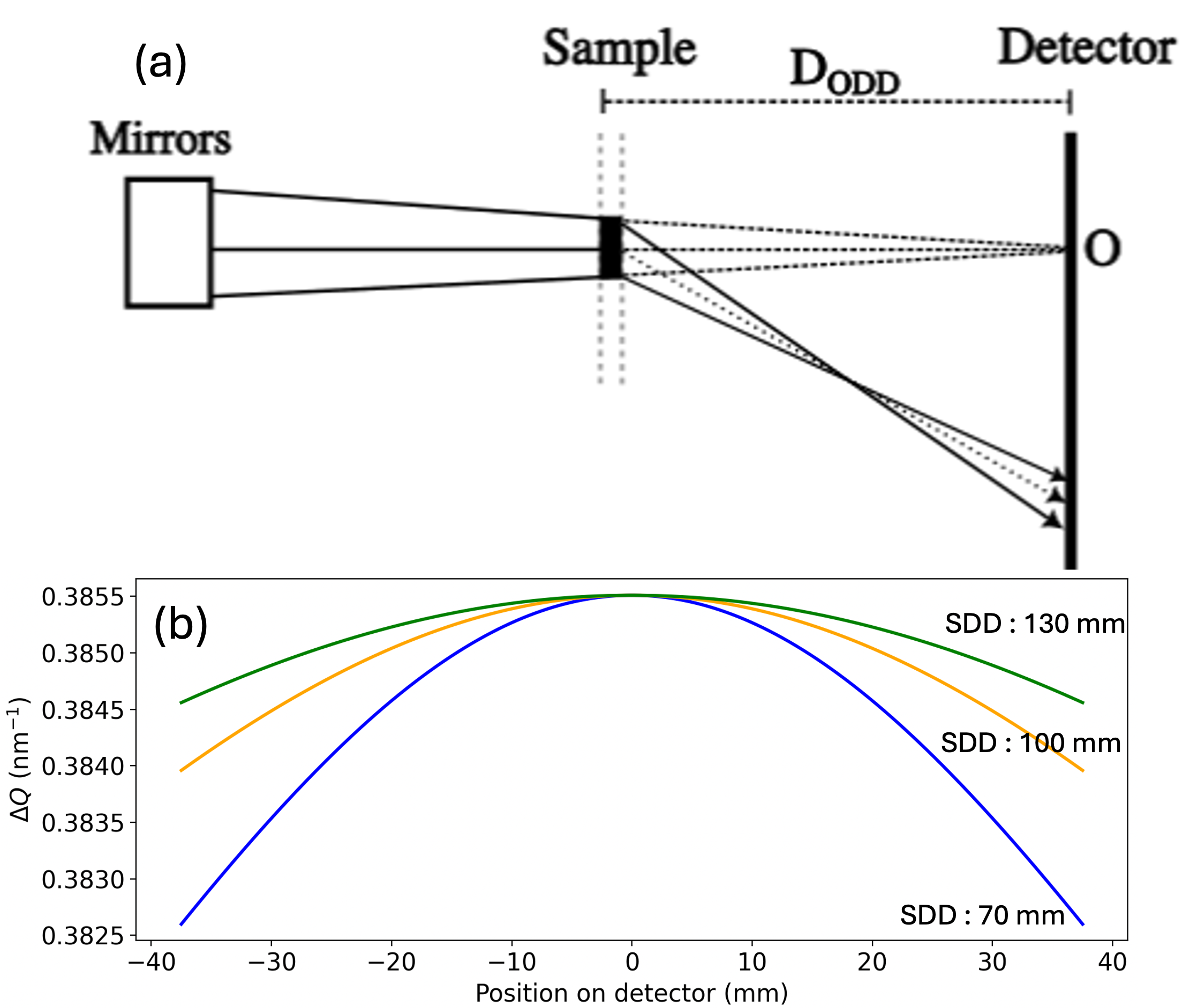}
    \caption{Effect of beam convergence/divergence on angular resolution at a $SDD$ of \qty{95}{\milli\meter}. The range of positions on this figure and those that follow correspond approximately to the size of the Eiger2 R 1M detector of MAXIMA.}
    \label{fig:beam divergence}
\end{figure}

In optimizing MAXIMA for throughput and spatial resolution, a tradeoff was made with respect to angular resolution. There are several factors that influence the angular resolution of a transmission XRD measurement. First, we consider effect of the beam divergence (actually, convergence) from the focusing mirrors, which is about \qty{3.15}{mrad}. This effect is shown in Fig.~\ref{fig:beam divergence}(a).

The effect of beam divergence on $q$-space resolution, represented as a top-hat function for simplicity, can be found from the definition of the magnitude of the scattering vector,
\begin{align}
Q               &=   \frac{4\pi}{\lambda}\sin\!\left(\theta\right)\\
\Delta Q_{divergence}              &=  \left(\frac{4\pi}{\lambda}\right) \cos\theta\ \Delta\theta
\label{scattering_vector_equation}
\end{align} where $\Delta Q$ is the angular broadening, $\lambda$ is the \xray source wavelength, $\theta$ is the diffraction angle, and $d\theta$ is the convergence angle. The range of $\Delta  Q$ caused by beam divergence at different positions on the detector is plotted in Fig.~\ref{fig:beam divergence}(b).

\begin{figure}[tbp] 
    \centering
\includegraphics[width=0.45\textwidth]{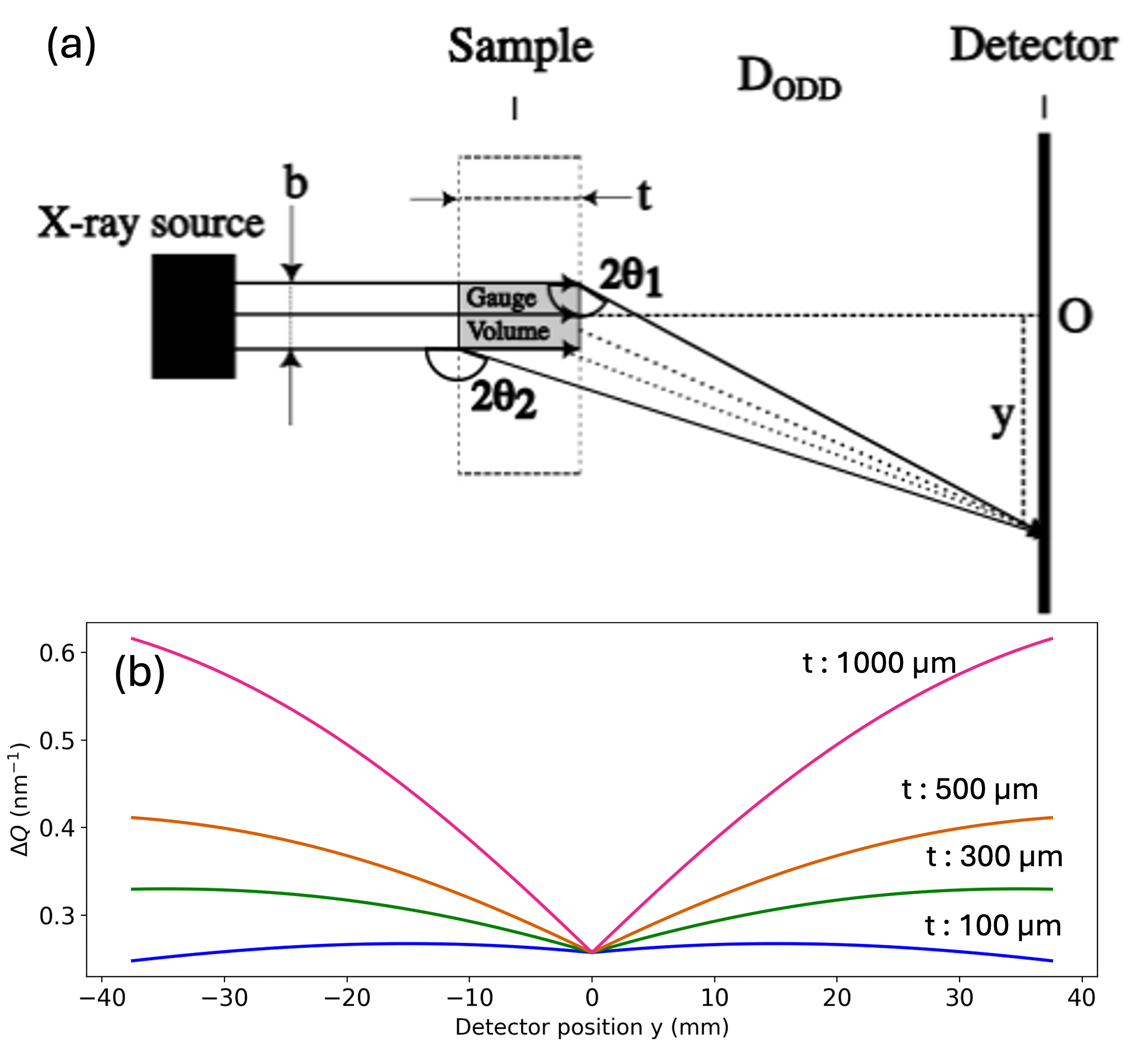}
    \caption{Effect of finite sample thickness and beam size on angular resolution at a $SDD$ of \qty{95}{\milli\meter} and a beam size of \qty{200}{\micro\meter}.}
    \label{fig:thickness broadening}
\end{figure}

A second contribution to instrumental comes from the effect of sample thickness and beam size. As shown in Fig.~\ref{fig:thickness broadening}(a), \xrays scattering from different regions of the scattering volume will strike a fixed point on the detector at different angles. This angular resolution is given by 
\begin{align}
\Delta Q_{\text{thickness}} &= \frac{4\pi}{\lambda} 
\Bigg( 
\sin\Big(\tan^{-1}\big(\frac{y + \frac{b}{2}}{SDD - \frac{t}{2}}\big)\Big) \nonumber\\
&\quad - 
\sin\Big(\tan^{-1}\big(\frac{y - \frac{b}{2}}{SDD + \frac{t}{2}}\big)\Big)
\Bigg)
\label{eq: thickness}
\end{align}

where $SDD$ is the sample-to-detector distance, $y$ is the distance from beam center, and $b$ is the width of the incident beam. These values are plotted as a function of position on the detector in Fig.~\ref{fig:thickness broadening}(b). Notice that for very thick samples ($\sim$\qty{1}{mm}) the loss of angular resolution is substantial, approaching that of the beam convergence discussed above.

A third effect on resolution is the range of angles spanned by a single pixel on the detector, which is given by 
\begin{equation}
\Delta Q_{\text{pixel size}} = \frac{4\pi}{\lambda} 
\sin\Bigg(\frac{SDD}{SDD^2 + y^2}\, \Delta y \Bigg)
\label{eqn:pixelsize}
\end{equation}
as described by He \cite{he_x-ray_2018}.
\begin{figure}[tbp]
    \centering
\includegraphics[width=0.4\textwidth]{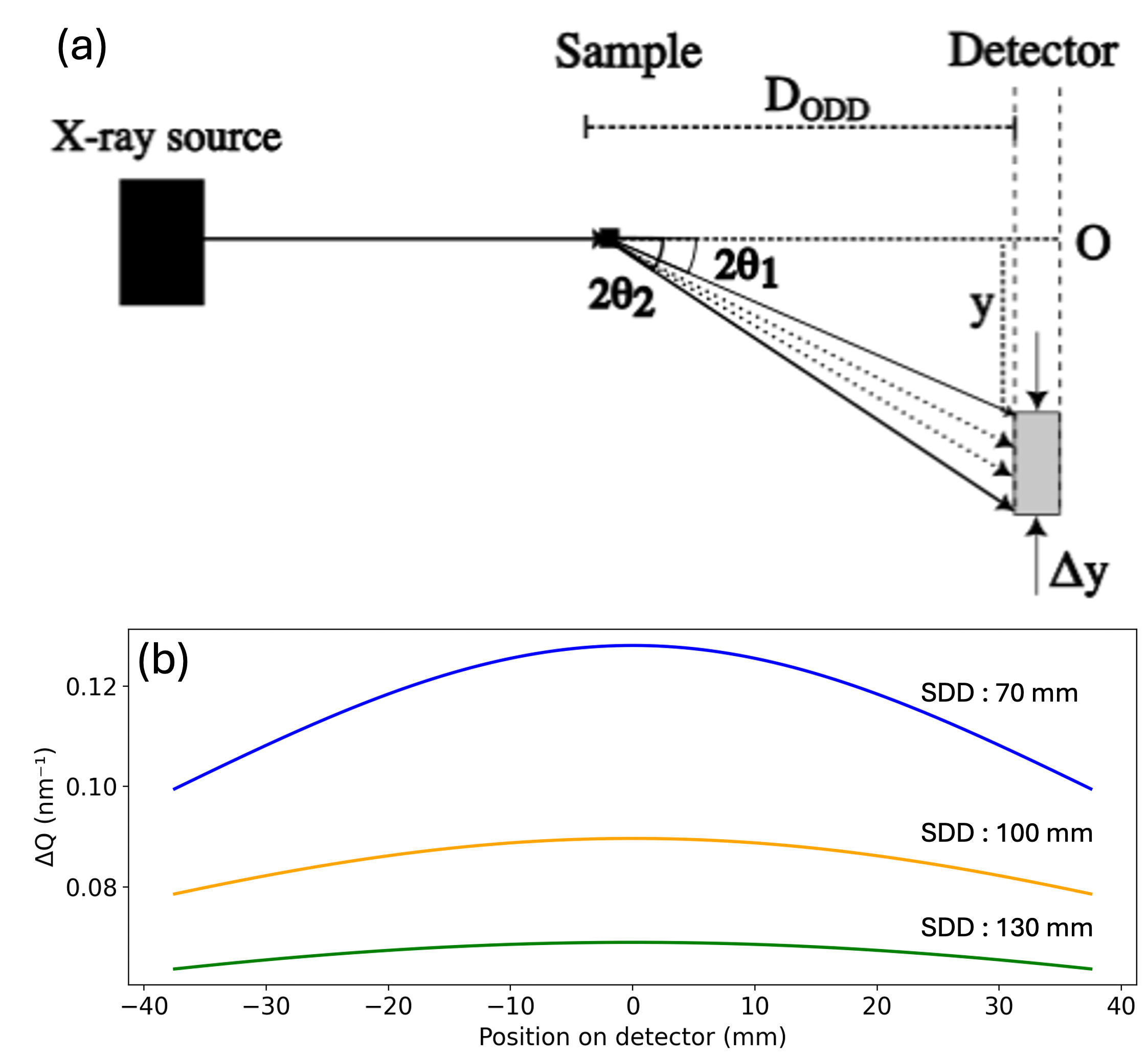}
    \caption{Effect of finite detector pixel size on angular resolution.}
    \label{fig:resolution}
\end{figure}
As shown in Fig.~\ref{fig:resolution}, this effect is comparable in magnitude to that of sample thickness and beam size (Fig.~\ref{fig:thickness broadening}) but smaller than the effect due to beam convergence (Fig.~\ref{fig:beam divergence}).

\begin{figure}[tbp] 
    \centering
    \includegraphics[width=0.45\textwidth]{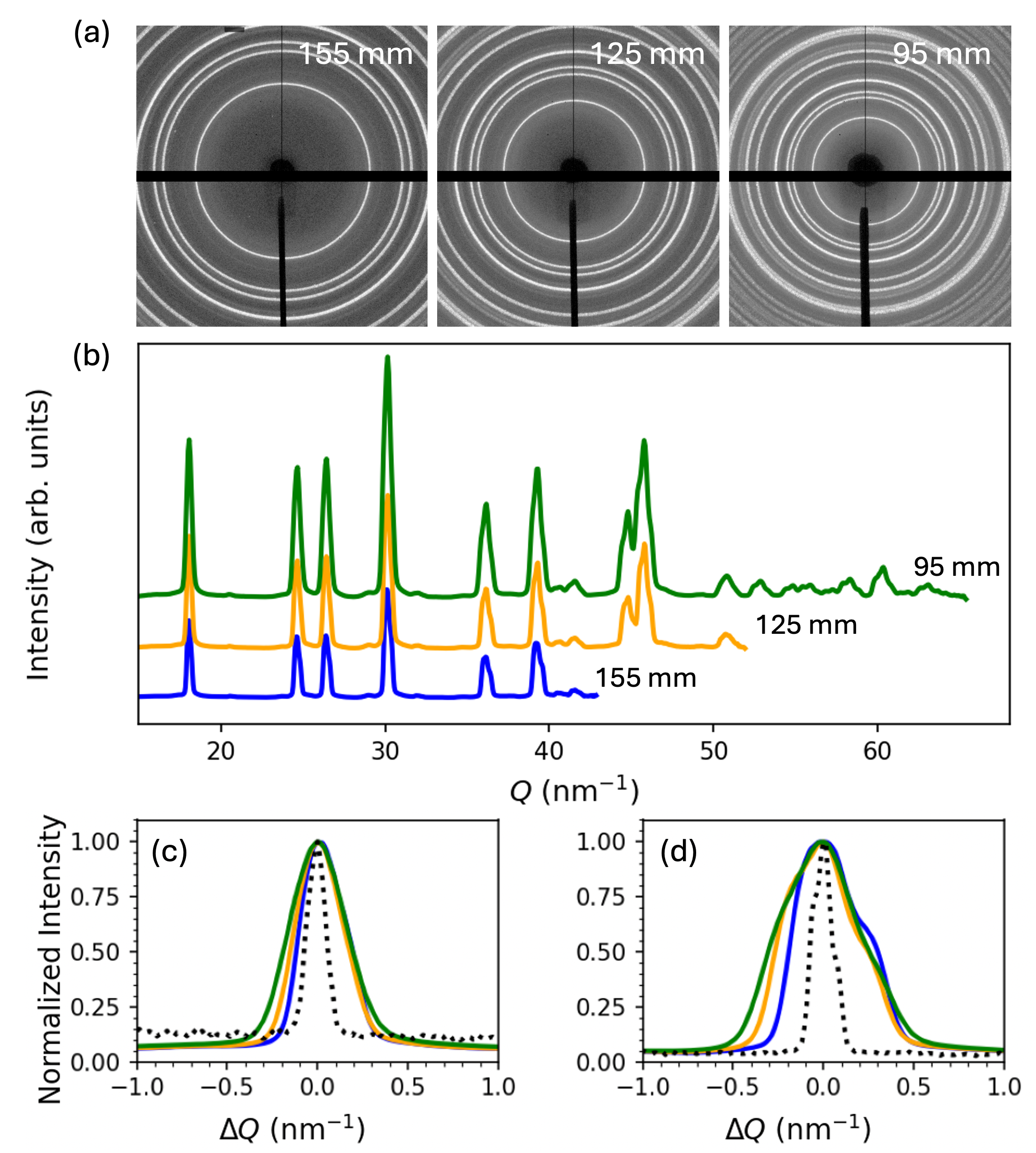}
    \caption{(a) 2D patterns and (b) 1D lineouts for NIST standard Al$_2$O$_3$ at three detector distances, along with azimuthally-integrated 1D patterns. Peaks for (c) (012) and (d) (113) are presented normalized with respect to intensity and centered around the maximum intensity value. These peaks are shown with comparison to instrumental broadening from a Bruker D8 Advance diffractometer scan (dotted black line) done with a \qty{1}{mm} divergence slit and a \qty{2.5}{\degree} Soller slit in a reflection geometry.
    }
    \label{fig:angular tradeoff}
\end{figure}

The discussion to this point has not separately considered the influence of the sample-to-detector distance ($SDD$ in Fig.~\ref{fig:beam divergence}). In MAXIMA this distance can be adjusted; increasing the detector distance improves resolution at the cost of reduced range of $Q$ covered by the XRD pattern. Figure~\ref{fig:angular tradeoff}(a) and (b) shows the increased $Q$ range achieved by moving the detector closer to the specimen. The corresponding loss of resolution is illustrated in Figure~\ref{fig:angular tradeoff}~(c) and (d). Those figures also show a comparison with a typical laboratory powder diffractometer, making clear that the high throughput of MAXIMA comes at the cost of angular resolution. It should be noted, however, that these data are from a \qty{1.5}{mm} thick standard, which is something of a worst case for a transmission diffractometer like MAXIMA. For more typical sample thickness (from \qtyrange{100}{500}{\micro\meter}) the effect on resolution is much smaller.

The overall angular resolution of the instrument is a convolution of all three of these contributions (plus other smaller effects not considered here). We calculate the overall resolution by adding the individual contributions in quadrature,
\begin{equation}
\Delta Q = \sqrt{\left(\Delta Q_\mathrm{divergence}\right)^2 +
\left(\Delta Q_\mathrm{thickness}\right)^2 +
\left(\Delta Q_\mathrm{pixel\ size}\right)^2
},
\end{equation}
where the individual contributions to $\Delta Q$ are taken from Eqns.~\ref{scattering_vector_equation}, \ref{eq: thickness}, and \ref{eqn:pixelsize}.

To estimate the angular broadening seen in our scans, we can convert this quadrature sum of $\Delta Q$ values (which we have implicitly treated as top-hat functions above) into the equivalent variance of a Gaussian profile by calculating the variance of a top-hat function \begin{equation}
\sigma_{tophat} =\frac{\Delta Q}{\sqrt{12}}
\label{sigma_conversion}
\end{equation} and calculating the full-width at half-max (FWHM) of a Gaussian with an equivalent second moment \begin{equation}
FWHM_{Equivalent} =2\sqrt{2\ln2}\sigma_{tophat}.
\label{Gaussian_fwhm}
\end{equation}

Table~\ref{tbl:summary} summarizes the contributions to resolution for the two alumina standard peaks in Fig.~\ref{fig:angular tradeoff}~(c) and (d).
\begin{table}[tb]
\centering
\begin{tabular}{lSS}
                                & \text{(012)} & \text{(113)} \\ \hline
$\Delta Q_\mathrm{divergence}$  & 0.38  & 0.38 \\
$\Delta Q_\mathrm{thickness}$   & 0.52  & 0.68 \\
$\Delta Q_\mathrm{pixel\ size}$ & 0.09  & 0.09  \\
$\Delta Q_\mathrm{total}$       & 0.66  & 0.79  \\
$FWHM_{Equivalent}$       & 0.45  & 0.53  \\

\end{tabular}

\caption{Contributions to resolution from beam divergence (Eqn.~\ref{scattering_vector_equation}), beam size and sample thickness (Eqn.~\ref{eq: thickness}), and pixel size (Eqn.~\ref{eqn:pixelsize}).}
\label{tbl:summary}
\end{table}
In both the (012) and (113) peaks, the calculated values slightly over-predict the peak widths of Fig.~\ref{fig:angular tradeoff}(c-d). We ascribe the difference to the simple way in which we have handled the convolution of the three contributions.

\section{High-throughput combinatorial study}
As a demonstration of the capabilities of MAXIMA for high-throughput studies, we performed a proof-of-concept experiment on combinatorial Cu-Ti alloy foils (Sec.~\ref{sec:performance}). We examined six graded specimens, with the most Cu-rich starting at nearly pure Cu and the most Ti-rich ending at about \qty{6}{at.\%} Ti. XRD and XRF data were recorded simultaneously at each of 27 points separated by \qty{1}{mm} on each sample, for a total of 162 scan points. As noted above, good XRD patterns can be collected in as little as \qty{1}{s}; in another recent demonstration we collected 1,500 diffraction patterns in about three hours. In the present experiment the data collection time was extended to \qty{30}{s} to ensure sufficient counting statistics for the Ti $K\alpha$ fluorescence. This is because the low energy of Ti $K\alpha$ (\qty{4.51}{keV}) means that it is strongly attenuated by the sample itself (absorption length of \qty{4.4}{\micro\meter}), so that only Ti near the surface can be detected. In addition, the \qty{5}{mm} of air between the sample and detector and \qty{0.5}{mm} beryllium window on the detector sequentially attenuate the signal by about \qty{6}{\%} and \qty{43}{\%}, respectively. Despite this relatively long time per point, the total data collection time for all 162 patterns, including overhead for transferring samples into/out of the instrument, was about \qty{90}{min}.

\begin{figure*}[htbp]
    \centering
    \includegraphics[width=\textwidth]{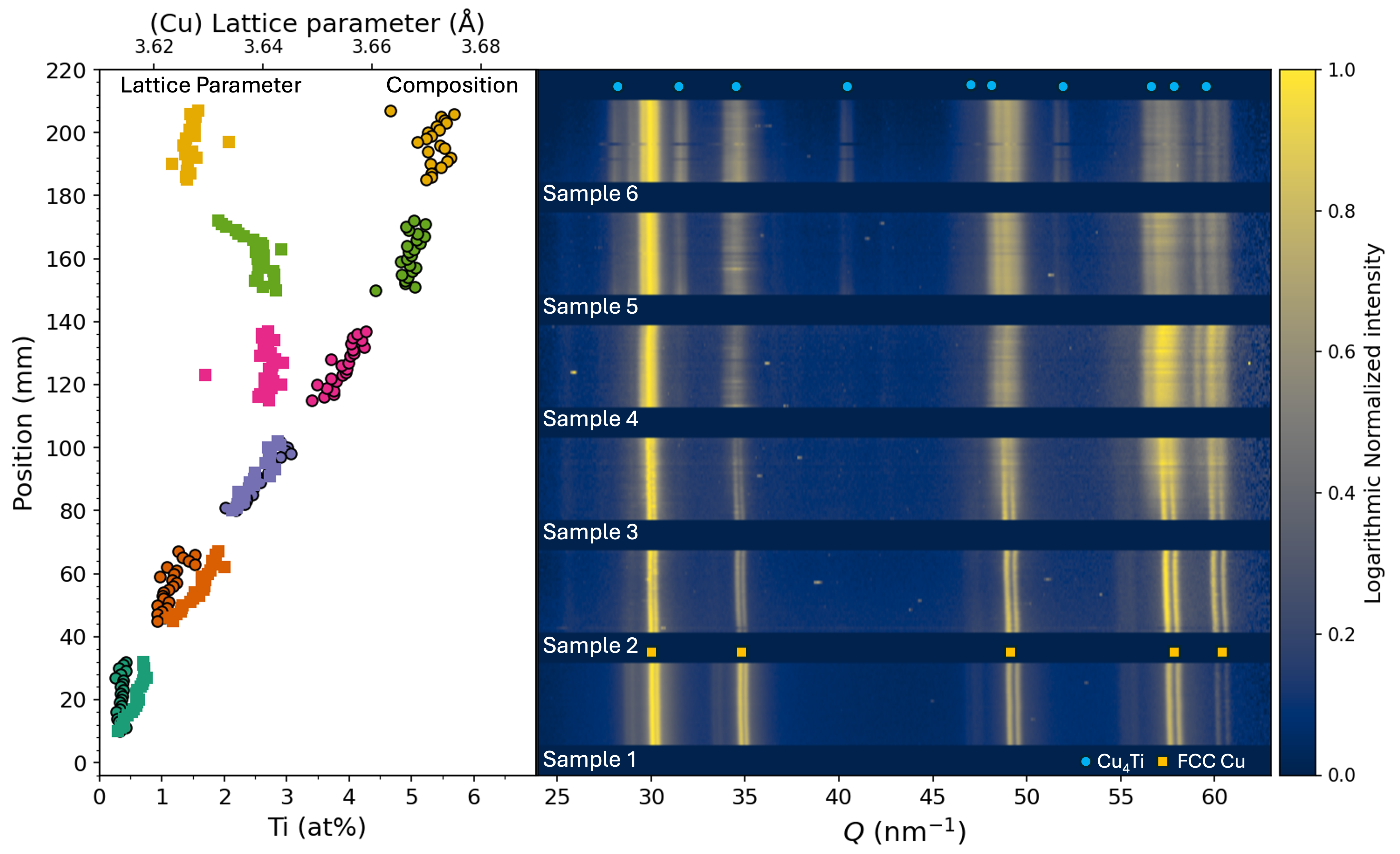}
    \caption{High-throughput characterization of six graded specimens of Cu-Ti alloys. At right is shown a heat map of the 162 XRD scans (27 per specimen), azimuthally integrated from the raw 2D patterns. (The gaps between the six sets of scans reflect physical gaps  between specimens on the substrate during vapor deposition. Note that the intensity map is on a logarithmic scale to highlight weak peaks.) At left is shown both the lattice parameter of the fcc Cu solid solution determined from the XRD peak positions and the elemental composition at each scan point determined from the XRF spectra (collected simultaneously with the XRD patterns). 
    }
    \label{fig:combinatorial_cuti}
\end{figure*}

The results of the combinatorial study are shown in Fig.~\ref{fig:combinatorial_cuti}. The Cu-rich samples (1-3) show only XRD from fcc Cu, with the Cu lattice parameter increasing with Ti content. At the opposite end, samples 5 and 6 have the most Ti and show clear evidence for the presence of the Cu$_4$Ti intermetallic. Another obvious difference is the width of the diffraction peaks, which are sharp for nearly pure Cu but conspicuously broadened for the most Ti-rich specimens. 

The behavior of the lattice parameter of the fcc Cu solid solution is interesting. We calculated these lattice parameters by plotting the spacing of each \textit{hkl} peak,
\begin{equation}
d_{hkl} = \frac{a}{\sqrt{h^2+k^2+l^2}}
\end{equation}
and finding the lattice parameter $a$ as the slope via linear regression. As shown in Fig.~\ref{fig:combinatorial_cuti} the measured lattice parameter at the Cu-rich end (Sample~1) is about \qty{3.615}{\angstrom}, in good agreement with values for pure Cu in the Inorganic Crystal Structure Database~\cite{finger_crystal_1978}. With increasing Ti content the Cu lattice parameter increases, reaching a maximum of \qty{3.64}{\angstrom} in Sample~4 at about \qty{4}{at.\% Ti}, in reasonably good agreement with prior results of Krull and Newman~\cite{krull_lattice_1970} and Nagarjuna and Sarma~\cite{nagarjuna_variation_1999}. With further increasing Ti content we see the lattice parameter fall, which coincides with the appearance of the Cu$_4$Ti intermetallic. Similar behavior was observed by Nagarjuna and Sarma, who associated it with the formation of the intermetallic during quenching following a solutionizing heat treatment~\cite{nagarjuna_variation_1999}. One benefit of large datasets from combinatorial samples such as these is that such trends are immediately clear; in Nagarjuna and Sarma~\cite{nagarjuna_variation_1999} there are only five data points.

\section{Discussion}
The lattice parameter variation mentioned above is only one example of information that can be derived from the XRD patterns. We are currently developing a workflow for automated identification of crystalline phases and quantitative phase fraction measurements based on the reference intensity ratio (RIR) method~\cite{li_calculating_2023}. Grain sizes can be estimated from the Scherrer equation~\cite{cullity_elements_2014} for small grains ($<\qty{50}{nm}$), with the Williamson-Hall method~\cite{cullity_elements_2014} in principle allowing determination of microstrain as well. But we note that MAXIMA was not designed for high resolution, and the relatively large instrumental broadening (Sec.~\ref{sec:resolution}) limits what can be learned from the peak widths. On the other hand, the distribution of intensity around the powder rings can reveal information about crystallographic texture, and for samples with large grains, diffraction spot counting can yield estimates of grain sizes and grain size distributions~\cite{he_miscellaneous_2018}. 

\section {Conclusions} 
In this paper we presented the design and implementation of MAXIMA, a multi-modal \xray  system designed for automated, high-throughput characterization of structural materials by XRD and XRF. By combining a high-brightness liquid metal anode source with focusing optics and a low-noise, high-sensitivity \xray\ detector we can collect hundreds of diffraction patterns per hour. The high energy of the In $K\alpha$ radiation (\qty{24}{keV}) allows measurement of true bulk (through-thickness) microstructure of thick samples, while the focusing optics provide spatial resolution ($\sim$\qty{250}{\micro\meter}) for studying small specimens or making spatially-resolved measurements on larger combinatorial specimens.

Sample handling, including sample transfer to/from the AIMD-L centralized conveyance, is fully automated. Data streaming off the instrument is also automated, as are workflows for basic data processing (calibration, detector corrections) for both XRD and XRF. By integrating MAXIMA into an automated data pipeline with reproducible, containerized analysis workflows, we ensure that the data produced is FAIR compliant and immediately useful for research teams, particularly those involved in training ML models. This reduces human intervention, accelerates feedback loops, and supports continuous model improvement in self-driving laboratory contexts. 

\section{Acknowledgments}
We gratefully acknowledge all of the other individuals involved in the AIMD-L project including, but certainly not limited to, Eric Walker, Matthew Shaeffer, and Joseph Nkansah-Mahaney.

This research was accomplished through projects sponsored by the Army Research Laboratory under Cooperative Agreement Numbers W911NF-22-2-0014, W911NF-23-2-0062, W911NF-22-2-0121, and W911NF-22-2-0101 (to DCE). The views and conclusions contained in this document are those of the authors and should not be interpreted as representing the official policies, either expressed or implied, of the Army Research Laboratory or the U.S. Government. The U.S. Government is authorized to reproduce and distribute reprints for Government purposes notwithstanding any copyright notation herein. 

Support for data and streaming was also provided by NSF-OAC 2129051 (to DCE). This work used Jetstream2 \cite{hancock_jetstream2_2021} at Indiana University through allocation MAT240117 from the Advanced Cyberinfrastructure Coordination Ecosystem: Services \& Support (ACCESS) program \cite{boerner_access_2023}, which is supported by National Science Foundation grants \#2138259, \#2138286, \#2138307, \#2137603, and \#2138296. Additional support for AIMD-L from the Johns Hopkins Whiting School of Engineering is gratefully acknowledged.

FAIR data supporting this work is available at \url{https://doi.org/10.34863/e8bx-pk70}.

\bibliography{references}

\end{document}